\documentclass{article}
\usepackage{amsmath}
\usepackage{amsfonts}
\usepackage[latin1]{inputenc}

\def\Tr {{\rm \;Tr \;}}

\newcommand {\pa}{\partial}

\newtheorem{lemma}{Lemma}[section]
\newtheorem{theorem}[lemma]{Theorem}
\newtheorem{proposition}[lemma]{Proposition}
\newtheorem{corollary}[lemma]{Corollary}
\newtheorem{remark}[lemma]{Remark}

\newtheorem{definition}[lemma]{Definition}

\def\beq{\begin{equation}}   \def\eeq{\end{equation}}
\def\bea{\begin{eqnarray}}  \def\eea{\end{eqnarray}}

\newcommand{\QED}{\mbox{}\hfill
\raisebox{-2pt}{\rule{5.6pt}{8pt}\rule{4pt}{0pt}}
            \medskip\par}
\newcommand\mysection{\setcounter{equation}{0}\section}
\renewcommand{\theequation}{\thesection.\arabic{equation}}
\newcounter{hran} \renewcommand{\thehran}{\thesection.\arabic{hran}}

\def\bmini{\setcounter{hran}{\value{equation}}
    \refstepcounter{hran}\setcounter{equation}{0}
    \renewcommand{\theequation}{\thehran\alph{equation}}\begin{eqnarray}}

\def\bminiG#1{\setcounter{hran}{\value{equation}}
\refstepcounter{hran}\setcounter{equation}{-1}
\renewcommand{\theequation}{\thehran\alph{equation}}
\refstepcounter{equation}\label{#1}\begin{eqnarray}}

\begin{document}

\title {Non Linear Eigenvalue Problems}
\author{ Didier Robert\\
D\'epartement de Math\'ematiques\\
Laboratoire Jean Leray, CNRS-UMR 6629\\
 Universit\'e de Nantes, 2 rue de la Houssini\`ere, \\
F-44322 NANTES Cedex 03, France\\
\it Didier.Robert@math.univ-nantes.fr}
\vskip 1 truecm
\date{}
\maketitle

\begin{abstract}
In this paper we consider generalized eigenvalue problems for a family of operators
 with a polynomial dependence on a complex parameter. This problem is equivalent to a genuine
  non self-adjoint operator. We discuss  here  existence of
   non trivial eigenstates for models coming from analytic theory of smoothness
    for P.D.E. We shall review some old results  and present recent improvements
     on this subject.

\end{abstract}

\pagestyle{myheadings}

\mysection{Introduction}
The problem considered in this paper has two very different origins. The first,  from the historical  point of view, 
 concerns {\em Dissipative Problems in Mechanics}. Let us consider the second order differential equation
 \beq\label{eq1}
 Au^{\prime\prime} + Bu^\prime + Cu = 0,
 \eeq
 where the unkonwn function $u$ is defined on $\mathbb R$ with values in some Hilbert space ${\cal H}$
  and $u^\prime = \dfrac{du}{dt}$. Equation \ref{eq1} is a model in mechanics for small oscillations of a continuum
   system in the presence of an impedence force \cite{krla}.\\
  Now looking for stationary solutions of  (\ref{eq1}), that means $u(t) = u_0{\rm e}^{\lambda t}$,
   we have the following equation
   \beq\label{eq2}
   (\lambda^2A + \lambda B + C)u_0= 0
   \eeq
   So equation (\ref{eq2}) is a non linear eigenvalue problem in the parameter $\lambda\in \mathbb C$.
  Existence of  non null solutions for (\ref{eq2}) is a non trivial problem. For $B\neq 0$ this problem
   is equivalent to a true non-selfajoint linear eigenvalue problem (see section II of this paper)
    and even the  existence of one solution for one complex number may be a difficult problem.
    But  by adding suitable  conditions on $A, B, C$ several authors, \cite{krla, ke, frsh, ma}
     have proved  the existence of a total set of generalized eigenfunctions for (\ref{eq2}).
     
     I  start working  on this subject 25 years ago for a completely different reason.  At that time 
      B. Helffer  asked   Pham The  Lai and me if the following non linear eigenvalue problem
       has at least a  solution ($\lambda, u$)  where $\lambda \in\mathbb C$
        and $u$ in the Schwartz space ${\cal S}(\mathbb R)$, $u\neq 0$, 
        \beq\label{eq3}
        \left(D_x^2 + (x^2 -\lambda)^2\right)u = 0,
        \eeq
        where $D_x = \dfrac{\partial}{i\partial x}$. 
        Equation (\ref{eq3}) is connected with analytic hypoellipticity for operators like sum of squares of
         analytic vector fields $X_1, \cdots, X_r$, defined  in an open set of
         $\mathbb R^n$,   satisfying the H\"ormander's condition:
          there exists an  integer $N$ such that the iterated brackets  of the fields $X_j$ of length less than $N$  
          span a vector space of dimension $n$ at each point. This is a sufficient condition for $C^\infty$-hypoellipticity
           for operator  $A = \sum X_j^2$ \cite{ho},  i.e  if $Au$ is $C^\infty$
           in the open set $\omega$ then $u$ is also  $C^\infty$ in the open set $\omega$.
           When the  coefficients of the $X_j$ are real-analytic in $\omega$,  satisfying  H\"ormander  condition,  and $Au$
            is  real-analytic in $\omega$, is it true  that $u$ is real-analytic in $\omega$?
            This is  the analytic-hypoellipticity problem. \\
            The general answer is no. The first example was given by Baouendi-Goulaouic \cite{bago} with the
             following system in  $\mathbb R^3$,
             \beq\label{eq4}
             X_1=\frac{\partial}{\partial x_1},\;X_2=\frac{\partial}{\partial x_2},\; X_3= x_1\frac{\partial}{\partial x_3}.
             \eeq
             For this example, Baouendi-Goulaouic  have constructed  a solution $u$, non analytic at 0, 
              such that $Au = 0$,  by using non trivial solutions
               of the equation $(D_{x_1}^2 + x_1^2 + \lambda^2)v = 0$
                which exist  for $\lambda = i\sqrt{2j+1}$, $j\in \mathbb N$,
                 as it is well known for harmonic oscillators. \\
                 In 1978, B. Helffer has proposed another example of sum of squares 
                  of vector fields  which are 
                  hypoelliptic but not analytic-hypoelliptic: 
                  $A = D_{x_1}^2 + \left(x_1^2D_{x_2} - D_{x_3}\right)^2$.
                   The  Baouendi-Goulaouic construction of  non analytic solutions at point $(0,0,0)$ for $A$
                  is also  possible if (\ref{eq3})  has a non trivial solution 
                  (for a generalization of this method see \cite{hahi1}).
               But this problem is less obvious than for harmonic oscillators.
                  In \cite{phro} we have given a positive answer to the question and  we have  
                             proved  furthermore  that
          there exists a total set of generalized eigenfunctions. Our proof uses  pseudodifferential technics
          (parametrices) and spectral analysis.
          Nowadays, two other proofs of this result are known.\\
        M. Christ \cite{ch1}, using  O.D.E techniques  and Wronskian arguments,  has extended our result
         to the equation
         \beq\label{eq4}
         \left(D_x^2 + (x^m -\lambda)^2\right)u = 0
         \eeq
          for every $m\in \mathbb N$,  $m\geq 2$.\\
          Let us remark that if  $m=1$ ,  for every $\lambda \in \mathbb C$, the equation  (\ref{eq4}) has 
           only the zero solution because of translation invariance.\\
        Recently Chanillo-Helffer-Laptev \cite{chhela} have given a proof using a very different
           and elegant method  involving  trace inequalities and  the Lidskii theorem concerning the trace
            of operators.  In the paper \cite{chhela} the authors consider  a more general problem,
            in several real  variables ($x\in\mathbb R^n$). 
             Let us introduce the following family of differential operators,
            \beq\label{eq5}
            L_P(\lambda) = -\triangle + \left(P(x) - \lambda\right)^2
            \eeq
            where $P$ is a  polynomial of degree $m\geq 2$ such that the homogeneous part $P_m$
             of $P$ satisfies $P_m(x) > 0$  for every $x\in \mathbb R^n\backslash\{0\}$
            (in other words we say that $P$ is a positive-elliptic polynomial). \\
            In \cite{chhela}  the authors prove existence of non trivial solutions for 
            $1\leq n\leq 3$  assuming that $m$  is large enough. \\
            In January 2003   Bernard Helffer  gave  in Nantes  a lecture 
             concerning the work  \cite{chhela}. After that,  Bernard, Xue Ping  (Wang) and me,  have  improved  in \cite{herowa}
               the results of  \cite{chhela} by making a  semi-classical analysis of the traces identities coming  from the Lidskii theorem.
                The main result  proved  in \cite{herowa} is the following
                \begin{theorem}
                Assume that $n$ is $\underline{even}$ and that $P$ is a positive-elliptic polynomial
                 of degree $m\geq 2$. \\
                 Then there exists  $\lambda\in \mathbb C$ and   $ u\in {\cal S}(\mathbb R^n)$, $u\neq 0$,
                  such that $L_P(\lambda)u = 0 $.
        \end{theorem}
        \begin{remark}
        Our proof gives an infinite number of solutions but it is not known,
          if the solutions of (\ref{eq4}) span the whole  Hilbert space
          $L^2(\mathbb R^n)$.\\
       On the other side for $n$ odd, $n\geq 3$,  the problem of existence of  non zero solutions is still open. 
       It seems reasonable to conjecture that such solutions always exist.
       It is  true for $n=1$ and for some cases if $n=3$.\\
       Another difficult but interesting problem would be to localize  in the complex plane
        these possible eigenvalues $\lambda$. We shall give  a very partial result at the end of this paper.
         \end{remark}

         In this paper we want to explain in more details   some results  concerning these non linear eigenvalue problems and to give the main steps
          of their  proofs.  We also explain an approach to prove the above conjecture in
           odd dimension, $n\geq 3$ (see also \cite{ne}).

           \section{Functional Analysis approach  of  the problem}
           We start here with a more general problem. Let be $k\in \mathbb N$,
           $n\geq 1$, and a pencil $L(\lambda)$ of operators
            in the Hilbert space ${\cal H} $ defined by
            \beq\label{eq5}
            L(\lambda) = H_0 + \lambda H_1 + \cdots + \lambda^{k-1}H_{k-1} + \lambda^k\mathbf 1
            \eeq
            Let us assume the following properties:\\
           (P-1) $H_0$ is a  self-adjoint, positive operator,  with domain $D(H_0)$ in ${\cal H}$.\\
           (P-2) For every $0\leq j\leq k-1$, $H_jH_0^{(j-k/)k}$
            and  $H_0^{(j-k)/k}H_j$ are bounded operators in ${\cal H}$.\\
            (P-3) $H_0^{-1/k}$ is in some Schatten class ${\cal C}^p$ for $p>0$ 
          ( for the definition of Schatten classes see \cite{gokr}).
          
          Then $L(\lambda)$, $\lambda\in \mathbb C$, is a family of closed operators on the domain
            $D(H_0)$. Moreover  the index of $L(\lambda)$ is 0 and
             $\lambda \mapsto L(\lambda)^{-1}$ is a meromorphic mapping from
              $\mathbb C$ into the Banach space ${\cal C(H)}$ of compact operators
               in ${\cal H}$. So $\lambda$ is a p\^ole for $L$ if and only if $L(\lambda)$
                is not injective  in $D(H_0)$. But according to a well known trick
                 the poles of $L$ can be identified with the eigenvalues of a non-self-adjoint
                  operator.  \\
                  Let us define the $k\times k$ matrix of operators
                  \beq\label{eqmat}
                  A_L = 
                  \left(
                  \begin{array}{ccccc}
                  0  &1  &0  &\ldots & 0\\
                  0 &0  &1 &\ldots &0\\
                  \vdots &0   &0 &\ddots& 0\\
                  0&0&0& \ldots & 1\\  
                  -H_0 & -H_1 & -H_2&\ldots& -H_{k-1}\\
                  \end{array}
                  \right).  
                  \eeq
           $A_L$ is a closed operator in the Hilbert space
           \beq
           {\cal K} = \prod_{1\leq j\leq k}D\left(H_0^{(k-j)/k}\right),
           \eeq
            with  domain
            \beq
            D(A_L) = \prod_{0\leq j\leq k-1}D\left(H_0^{(k-j)/k}\right).
            \eeq   
            $A_L$  is  invertible and    $A_L^{-1}$  is in the Schatten class ${\cal C}^p$.
             For $ \lambda \in \mathbb C$, $\lambda \neq 0$, we have
             \beq
            \{L(\lambda)\;  {\rm is \; invertible } \} \Longleftrightarrow   \{A_L - \lambda\mathbf 1\; {\rm is \;\; invertible} \}
            \eeq
            
            Moreover, if we write down the resolvent of $A_L$ as a matrix operator, 
            \beq
            (A_L - \lambda\mathbf 1)^{-1} =
             \{r_{j,\ell}(\lambda)\}_{0\leq j, \leq \ell\leq k-1}
            \eeq
            then  we have $L(\lambda)^{-1} = -r_{0,k-1}(\lambda)$.\\
            Let us denote by ${\rm sp}[L]$ the set of eigenvalues of $L$ and if $\lambda_0\in {\rm sp}[L]$,
             ${\cal E}_{\lambda_0}[L]$ denotes the generalized eigenspace for the eigenvalue $\lambda_0$,
          defined  by  Keldysh \cite{ke},    as the linear space span by the solutions  
                 $u_0, u_1, \cdots, u_\ell$ of the following system of equations ($\ell \in \mathbb N$),
                \bea
                L(\lambda_0)u_0 & = &  0\\
                 L(\lambda_0)u_\ell + \frac{ dL(\lambda_0)}{d\lambda}u_{\ell -1} + \cdots
                  +  \frac{ d^\ell L(\lambda_0)}{d\lambda^\ell}u_0 & = & 0.
                  \eea          
            The following result is proved in \cite{ke, phro}
            \begin{lemma}
            If the linear space $\displaystyle{\bigoplus_{\lambda\in\mathbb C}{\cal E}_\lambda[A_L]}$  is dense in ${\cal H}$
             then $\displaystyle{\bigoplus_{\lambda\in\mathbb C}{\cal E}_\lambda[L]} $   is dense in ${\cal K}$.
             \end{lemma}
             
             This lemma is useful  because we can apply   known results for non self-adjoint operator 
              to our non linear eigenvalues problem.
              \begin{theorem}[Dunford-Schwartz, \cite{dusc}] Assume that there exist rays $\Xi_j$, $1\leq j\leq J$
               in the complex plane $\mathbb C$, starting from the origin, such that the angle between two consecutive rays is strictly smaller than $\pi/p$ and there exist  positive real numbers $\rho, R$,  such that
               \beq
               \Vert L(\lambda)\Vert_{\cal L(H)} = {\cal O}(\vert\lambda\vert^\rho), \;\;
                {\rm for}\;\;  \vert\lambda\vert \geq R, \;\;
                 \lambda \in \cup_{1\leq j\leq J} \Xi_j.
                 \eeq
              Then  $\displaystyle{\bigoplus_{\lambda\in\mathbb C}{\cal E}_\lambda[L]}$  is dense in ${\cal H}$.
              \end{theorem}
             Following \cite{phro}, we can apply the above functional analysis result to quadratic pencils
              $L_m(\lambda) = D_x^2 + (x^m -\lambda)^2$, for $m\in\mathbb N$,  $m$ even, $m \geq 2$.
              \begin{proposition}\label{base}
              $L_m(\lambda)$ has a complete system of generalized eigenfunctions in $L^2(\mathbb R)$.
               These functions are in the Schwartz space ${\cal S}(\mathbb R)$.
               \end{proposition}
               {\bf Sketch of Proof}:\\
               We have here  $L(\lambda) = D_x^2 + x^{2m} -2\lambda x^m + \lambda^2$
                and $H_0^{-1/2}$ is in ${\cal C}^p$ for every $p > \dfrac{m+1}{m}$.  
                 So the angle condition of the Dunford-Schwartz Theorem  is here 
                 $\theta < \dfrac{m\pi}{m+1}$.
                 Let us denote $\Xi_\alpha = \{ r{\rm e}^{i\alpha}, r\geq 0\}$. Then proposition  \ref{base}
                 is a consequence of  the following  result:
                  \begin{lemma}
                  $L(\lambda)^{-1}$ exists on $\Xi_0$ and on  $\Xi_\alpha$
                   for every
                    $\alpha \in ]\pi/2, \pi]\cup[-\pi, -\pi/2[$ and satisfies the estimates
                    \bea
                    \Vert L(\lambda)^{-1}\Vert = {\cal O}\left(\vert\lambda\vert^{1/2m}\right) {\rm for}\; \lambda \in \Xi_0,\\
                     \Vert L(\lambda)^{-1}\Vert = {\cal O}\left(\vert\lambda\vert^{-2}\right) {\rm for}\; \lambda \in \Xi_\alpha, \;
                     \alpha \in ]\pi/2, \pi]\cup[-\pi, -\pi/2[.
                     \eea
                  \end{lemma}
                   {\bf Sketch of Proof  of the Lemma 2.4}\\
                   The estimate on  $\Xi_0$ comes from a direct computation.
                   Estimate on $\Xi_\alpha$ can be proved by pseudodifferential technics (cf \cite{phro}) or also by direct estimates. \QED

                  The angle condition is more difficult to check in higher dimension $n$.
                   The reason is the following.  
                   For $L_P(\lambda) = -\triangle +(P(x)-\lambda)^2$  we have
                    $H_0 = -\triangle + P^2(x)$, where $P(x)$ is like $\vert x\vert^m$. Because of eigenvalue asymptotics
                     (see for exemple \cite{ro1}) we have 
                      $H_0^{-1/2}$ is in ${\cal C}^p$ for every $p > \dfrac{n(m+1)}{m}$,
                       this is optimal,  so  the angle condition to apply the 
                        Dunford-Schwartz theorem is 
                        $\theta < \dfrac{m\pi}{n(m+1)}$. Then  we need resolvent estimates
                         on closer and closer rays when $n$ increases.\\
                          It is  the reason why the approach proposed by 
                           Chanillo-Helffer-Laptev is very useful. The basic idea is the following.
                          Let us recall the Lidskii Theorem \cite{gokr}.
                          Let ${\cal H}$  be an Hilbert space and  $T$ an operator  in ${\cal H}$
                        in the Schatten class ${\cal C}^{1}$.  Then we have
                            \beq\label{eq6}
                            \Tr(T) = \sum_{\lambda\in {\rm sp}[T]} \lambda
                            \eeq
                             where ${\rm sp}[T]$ is the set of eigenvalues of $T$,
                              with their   multiplicities. \\
                              So to prove that $L(\lambda)$ has a non empty spectrum  
                                it is sufficient to prove     that
                               \beq\label{eq7}
                               \Tr\!\left(A_L\right)^{-\ell} \neq 0\;\; {\rm for\;\;some} \;\;\ell \geq p
                               \eeq
                       The property (\ref{eq7}) is the core of the method of \cite{chhela}.
                       For simplicity let us explain  this method in more details  in the one dimension case,
                        for $L_m$ ($m \geq 2$). For the computations
                         it is more convenient to conjugate  $A_L$  by a unitary operator
                          such that we get an operator $\tilde A_L$      acting in
                           the Hilbert space $L^2(\mathbb R)\times L^2(\mathbb R)$.
                           For $L = L_m$  we denote $A_m = \tilde A_{L_m}$
                            and we easily have
                            \beq
                            A_m = \left(
                            \begin{array}{cc}
                            0 & H_0^{1/2}\\
                            -H_0^{1/2}&-H_1\\
                            \end{array}
                            \right),
                            \eeq
                            where $H_0 = D_x^2 + x^{2m}$ and $H_1 = -2x^m$.\\
                             So we have
                             \beq
                            A_m^{-2} = \left(
                            \begin{array}{cc}
                          B^2-C^2& -BC\\
                          CB&-C^2\\
                            \end{array}
                            \right)
                            \eeq

                   where $C =  H_0^{-1/2} $,  $B= -H_0^{-1/2}H_1H_0^{-1/2}$.  Hence
                    we have
                    \beq\label{eq8}
                    \Tr\left(A_m^{-2}\right) = \Tr\left(B^2 - 2C^2\right).
                    \eeq
                          
                          By scaling we have, for every $\gamma>0$,
                          \beq
                          \Tr\left(D_x^2+\gamma x^{2m}\right)^{-1} =
                           \gamma^{1/(m+1)}\Tr\left(D_x^2+\gamma x^{2m}\right)^{-1}.
                           \eeq
                           Then, 
            computing the derivative at $\gamma = 1$ on each side, using the 
           Cauchy-Schwarz  inequality,  $\vert\Tr(M^2)\vert \leq \Tr(MM^*)$ for  $M = H_0^{-1}H_1$, we get
             \beq 
             \Tr(B^2) \leq \frac{4}{m+1}\Tr(C^2).
             \eeq
             So the conclusion follows, for  $m\geq 2$, with
             \beq 
               \Tr\!\left(A_m^{-2}\right) \leq \left(\frac{4}{m+1} -2\right)
               \Tr\left(H_0^{-1}\right)  <  0.
               \eeq
               In \cite{chhela} the authors have used the same method for $n = 2, 3$,
                by computing  $\Tr\left(A_L^{-4}\right)$. The proof is much more tricky
                 and give the expected conclusion for $m\geq 6$. 
                 We shall see in  the next
                  section that  by adding some semiclassical ingredients  in the Chanillo-Helffer-Laptev  approach as we did in \cite{herowa},
                    it is possible to improve their result in the even dimension case.

     \section{A Semiclassical Analysis of the problem}
    Let us consider first the quadratic pencil $L_P$ where $P$ is a positive-elliptic
     polynomial of degree $m\geq 2$ in $\mathbb R^n$, $n\geq 2$. For simplicity
      we assume that $P$ is homogeneous.   By the  scaling transformation 
      $x = \tau y$ with $\hbar = \tau^{1-m}$ and 
      $\mu = \dfrac{\lambda}{\tau^m} +1$ we  can  see that $L_P(\lambda)$
       is unitary equivalent to the semiclassical Hamiltonian
        $\hat H(\mu)$ where 
        \beq\label{eq9}
        \hat H(\mu) = -\hbar^2\triangle_y + \left(P(y) +1 -\mu\right)^2
        \eeq
        So $ \hat H(\mu) $ is the $\hbar$-Weyl operator with the symbol
         $H(\mu, y, \eta) = \eta^2 +  \left(P(y) + 1 -\mu\right)^2$. 
        For semiclassical analysis tools and  $\hbar$- Weyl quantization we refer to \cite{ro2} for scalar symbols and to  \cite{ko} for matrix symbols.
         Here we use the notation $\hat H$ for the  $\hbar$-Weyl quantization
         of the symbol $H$.
        \\
        As above, to the semiclassical quadratic pencil  $ \hat H(\mu) $ is associated
          a non self-adjoint matricial operator $\hat A_P$ in 
          $L^2(\mathbb R^n)\times L^2(\mathbb R^n)$,
           \beq
                            \hat A_P = \left(
                            \begin{array}{cc}
                            0 & \hat H_0^{1/2}\\
                            -\hat H_0^{1/2}&-\hat H_1\\
                            \end{array}
                            \right)
                            \eeq
        where $\hat H_0 =   -\hbar^2\triangle_y + \left(P(y) +1\right)^2$.
       The $\hbar$-symbol $A_P(y, \eta)$ of $\hat A_P$ has two eigenvalues
        \beq 
        \mu_\pm(y, \eta) = P(y) + 1 \pm i\vert\eta\vert
       \eeq
        So using standard methods in semiclassical and spectral analysis
     adapted from R.  Seeley \cite{se} and \cite{ro1}, the authors  get the following result
     \begin{theorem}\label{trace}
     For every real number $s < -\dfrac{n(m+1)}{m}$, in the semiclassical  regime
       $\hbar \searrow 0$, we have,
       \beq\label{eq10}
       \Tr\left(\hat A_P^s\right)  \asymp \sum_{j\geq 0}c_{j,s}\hbar^{j-n}
       \eeq
       with 
       \bea
       c_{0,s} &  =  & (2\pi)^{-n}\int_{\mathbb R^{2n}}\Re\left(\mu_+(x,\xi)\right)^sdxd\xi,\\
       c_{1,s} &  =  & 0, \\
       c_{2,s} &  = &  \; {\rm << something\;\; computable >>}
       \eea
       \end{theorem}
       So  using  Lidskii Theorem  to prove that $L_P(\lambda)$ has a non empty
        spectrum, it is enough to prove that there exist $s < -\dfrac{n(m+1)}{m}$
         and $j\in \mathbb N$ such that $c_{j,s} \neq 0$. The main result in 
       \cite{herowa} is that this can be checked if $n$ is even.
       \begin{lemma}\label{even}
       If $m\geq 2$ and $n$ is even then  for every $s < -\dfrac{n(m+1)}{m}$,
         $c_{0,s} \neq 0$.
              \end{lemma}
              {\bf Proof}\\
              We have
              \beq
              \int_{\mathbb R^{2n}}\mu_+(x,\xi)^sdxd\xi =
               \int_{\mathbb R^n}\left(P(x) + 1\right)^{s+n}\int_{\mathbb R^n}
               (1 + i\vert\eta\vert)^sd\eta
               \eeq
               So we have to compute $f_s(i)$ where
               \beq
               f_s(\alpha) = \int_{\mathbb R^n}(1 + \alpha\vert\eta\vert)^sd\eta.
               \eeq
               By scaling and analytic extension we easily get
               $ f_s(\alpha) = \alpha^{-n}f(1)$ for $\alpha \in \mathbb C$, 
               $ \alpha \notin ]-\infty, 0]$. 
               Then $\Re f_s(i) = \cos\left(\dfrac{n\pi}{2}\right)f_s(1)$ and $f_s(1) > 0$.
                So the conclusion follows. \QED
                \begin{remark}
                We conjecture that for $n$ odd, $n\geq 3$,  there always exists  $j\geq 1$
                 and $s< -\dfrac{n(m+1)}{m}$ such that $c_{j,s}\neq 0$.  To check this it is necessary
                  to perform  algebraic computations 
                  which are under investigation in \cite{ab}.  A similar method was used in \cite{ne}
                   to prove existence of resonances for matrix Schr\"odinger operators.
                     \end{remark}
                The above lemma gives  much more than existence of at least one  
                 eigenstate.  We shall see  that  there exists an infinite number of eigenstates and  give 
                  an estimate of their density.  It is known that  existence of resonances can be proved
                   as a consequence of a trace formula \cite{Sj, ne}. The same method can be applied
                    here, in an easier way, to estimate the number of eigenvalues.
                   Let us write the  generalized eingenvalues
                   $\{\lambda_j\}_{j\geq 1}$
                   of $L_P$  by increasing order of their modulus, repeated according their mutiplicities.
                    Let us introduce the counting  function 
                    \beq
                    N_L(r) = \#\{j\geq 1,\; \vert\lambda_j\vert \leq r \}.
                    \eeq
                  \begin{proposition}\label{estimate}
                 Under the assumption of Theorem  \ref{trace} and Lemma \ref{even}, 
                  there exists a constant $C > 0$ such that for every $r\geq 1$ we have
                 \beq
                  \frac{r^{n(m+1)/m}}{C} \leq N_L(r) \leq  Cr^{n(m+1)/m}
                   \eeq
                  
                  \end{proposition}  
                  {\bf Proof}\\
                  Let us denote $\theta  = \dfrac{n(m+1)}{m}$  and fix an integer $k > \theta$.
                   By a change of parameter, it results from Theorem \ref{trace} that we have
                   \beq
                   \sum_{j \geq 1}  (t+ \lambda_j)^{-k}  = c_0t^{\theta-k} + {\cal O}\left(t^{\theta - k -(m+1)/m}\right)
                   \eeq
                       where $c_0 \neq 0$.\\
                     To find an upper bound we apply the Weyl-Ky Fan inequality \cite{gokr}
                     \beq
                     \sum_{j \geq 1} (t + \vert\lambda_j\vert)^{-k }\leq 
                        \sum_{j \geq 1} (t + s_j)^{-k}
                        \eeq
                         where $\{s_j\}_{j\geq 1}$ is the set  of eigenvalues of $(A_L^*A_L)^{1/2}$.
                         But we also have a trace formula for  $(A_L^*A_L)^{1/2}$,  so we have
                         \beq
                           \sum_{j\geq 1} (t + \vert\lambda_j\vert)^{-k}  = {\cal O}(t^{\theta -k})
                           \eeq
                           which gives easily (taking $ t=r$) the upper bound:
                           \beq
                             N_L(r) \leq  Cr^{n(m+1)/m}
                           \eeq
                           For the lower bound, we first remark that  for $t\geq t_0$, $t_0$ large enough,
                            we have $\vert t+\lambda_j\vert \geq \dfrac{\vert\lambda_j\vert +t}{8}$.
                            This is true because,  for every $\varepsilon > 0$,  we have 
                             $\arg(\lambda_j)\in [-\pi/2-\varepsilon, \pi/2+\varepsilon]$
                              for $j$ large enough \cite{phro}.  Then  there exists $c_1> 0$ such that
                              \beq
                                \sum_{j \geq 1}(t + \vert\lambda_j\vert)^{-k }  \geq        
                               c_1 t^{\theta - k}
                              \eeq
                           It is convenient to write the above inequality with Stielj\`es integral
                           \beq
                           \int_0^\infty (t+r)^{-k}dN_L(r) \geq c_1t^{\theta-k}
                           \eeq
                           Let $\gamma > 1$  be a large constant to be chosen later.
                           We have, using the upper bound and an  integration by part,
                           \beq
                             \int_{\gamma t}^\infty (t+r)^{-k}dN_L(r)
                             \leq kC \left(\int_{\gamma}^\infty (1+u)^{-k-1}u^\theta du\right)t^{\theta-k}
                             \eeq
                            So we can choose $\gamma $ large enough such that
                            \beq
                             \int_0^{\gamma t }(t+r)^{-k}dN_L(r)\geq \frac{c_1}{2}t^{\theta-k}
                             \eeq
                             which easily gives the lower bound:
                             \beq
                             N_L(\gamma t) \geq    \frac{c_1}{2}t^\theta.
                             \eeq
                             \QED
                             In the paper \cite{herowa} we also consider the following quadratic pencils
                             \beq
                             L_{P,Q}(\lambda) = -\triangle + (P(x) -\lambda)^2 + Q(x)^2
                             \eeq
                              where  we assume that $P, Q$ are homogeneous polynomials of degree $m\geq 2$,
                               $P\geq 0$, $P^2+Q^2$ is elliptic and $Q$ {\bf is not identically 0} if $n$ is odd.
                                Then we can extend  Proposition \ref{estimate} to the corresponding counting function  $N_{ L_{P,Q}}(r)$. 
                                 \begin{proposition}
                                 The following estimates  are satisfied
                                  \beq
                  \frac{1}{C}r^{n(m+1)/m} \leq N_{L_{P,Q}}(r) \leq  Cr^{n(m+1)/m},
                   \eeq
                   if one of the following condition is satisfied\\
                   (i) $m\geq 2$, $n =1, 3$,\\
                   (ii) $n= 2$, $m\geq 3$,\\
                   (iii) $n=2$, $m=3$ and the following technical condition
                   \beq
                   \frac{(P+1)^2}{(P+1)^2+Q^2} -  \frac{3-\sqrt 2}{4}
                   \eeq
                 is  everywhere non negative, or everywhere non positive,  on $\mathbb R^n$.
                      \end{proposition} 
                      {\bf Sketch of Proof} (see \cite{herowa} for details).\\
                     We prove  that the leading coefficient $c_{0,s}$ in the trace formula is not 0.
                      \begin{remark}
                      The same results holds if $P, Q$ are polynomials such that the assumptions are  satisfied    
                               for     their homogeneous part of degree $m$. Furthermore we can replace the homogeneity condition
                        by a quasi-homogeneity condition like in the example
                         $P(x_1, x_2) = x_1^2  + x_2^4$.
                       
                      \end{remark}

                            \section{Localization of some eigenvalues}
                            We revisit here an example coming from a question that  G. M\'etivier  asked me twenty years ago.
                            Let us consider  the following quadratic pencil depending on  a  large parameter 
                            $\eta > 0$.
                            \beq
                            L_\eta(\lambda) = -\triangle_x + (P(x)-\lambda)^2 + \eta^2.
                            \eeq
                             Assuming as before that  $P$ is an elliptic-positive homogeneous polynomial of degree
                              $m\geq 2$. $  L_\eta(\lambda)$  
                              is conjugate to a semiclassical  operator  where    
                               $\hbar = \eta^{-(m+1)/m}$  and  $\mu = \dfrac{\lambda}{\eta}$.  More precisely,
                                we have
                                \beq
                                L_\eta(\lambda) = \eta^2\left(-\hbar^2\triangle_y + (P(y) -\mu)^2 + 1\right).
                                \eeq
                                The analogue of the trace Theorem \ref{trace} gives here for the leading coefficient
                                 \beq
                                     c_{0,s} = (2\pi)^{-n}\int_{\mathbb R^{2n}}2\Re\left(P(x) +  i\sqrt{1+\xi^2}\right)^s
                                     \eeq
                                     A direct computation, as we  did  in the proof of  Lemma \ref{even}, gives
                                     \beq
                                      c_{s,o} = \gamma_s\cos\left(\frac{(n+sm)\pi}{2m}\right),
                                      \eeq
                                      where $\gamma_s\neq 0$. So, for every $n\geq 1$,  there exists $s < -\dfrac{n(m+1)}{m}$, 
                                       such that $ c_{0,s}  \neq 0$.  So we get
                                  \begin{proposition}
                                  There exists $\eta_0 > 0$,  large enough,  such that,  for every $\eta \geq \eta_0$,
                                    $ L_\eta(\lambda)$ has a non empty spectrum.
                                    \end{proposition}
                                                     Let  $\{\lambda_j(\eta)\}_{j\geq 1}$  be the sequence  of eigenvalues of $L_\eta$, 
                                                        ordered by increasing modulus. So $\lambda_1(\eta)$ is a generalized eigenvalue
                                      of minimal modulus.  The question of  G. M\'etivier  was about the behaviour
                                       of $\lambda_1(\eta)$ as $\eta \rightarrow +\infty$.
                                       The answer is
                                       \begin{proposition}\label{loc}\cite{ro3}
                                       \beq
                                       \lim_{\eta \nearrow +\infty}\frac{\lambda_1(\eta)}{\eta} = \pm i
                                       \eeq
                                     (We have $\pm i$ because  $\lambda$  is eigenvalue if and only if   $\bar\lambda$ is
                                      eigenvalue).
                                      \end{proposition}
                                      {\bf Sketch of the  Proof}\\
                                      On the semiclassical side we have to prove
                                      \beq
                                      \lim_{\hbar\searrow 0}\mu_0(\hbar) = \pm i
                                      \eeq
                                      where  $ \mu_0(\hbar)$ is an eigenvalue of minimal modulus
                                       of  $\hat L(\mu) = -\hbar^2\triangle_y + (P(y) -\mu)^2 + 1$.
                                        If $\hat A $  is the non self-adjoint matricial operator associated with
                                         $  \hat L(\mu)$, let us introduce the family of complex variable functions
                                         \beq
                                         F_\hbar(z) = (2\pi\hbar)^n\Tr\!\left(\hat A - z\right)^{-N}
                                         \eeq
                                          where $N$ is chosen large enough. Let us denote 
                                          $\Omega_r =  \{ z\in \mathbb C,\;\; \vert z\vert < r\}$.
                                          The $\hbar$ principal symbol of $\hat A$
                                           has the eigenvalues $\mu_\pm(x,\xi) = P(x) \pm i\sqrt{1+\xi^2}  $.
                                            So, by standard  parametrix construction, for every $\alpha > 0$,
                                                    there exits $\varepsilon>0$
                                             such that for $\hbar < \varepsilon$,   $ F_\hbar(z)$ is holomorphic
                                              in  the set  
                                              $ B_\alpha = \{z\in\mathbb C,\;\; \Re z<1-\alpha\}\cup \{z\in\mathbb C,\;\;\vert\Im z\vert < 1-\alpha\}$.
                                              In particular   $ F_\hbar(z)$ is holomorphic in 
                                                              $\Omega_r := \{z\in \mathbb C, \vert z\vert z < r\}$  
                                                               for  $r < 1$, and  we have
                                              \beq
                                              \lim_{\hbar\searrow 0}F_\hbar(z) = F_{c\ell}(z)
                                              \eeq
                                               where
                                               \beq
                                                F_{c\ell}(z) = \int_{\mathbb R^{2n}}[\left(\mu_+(x,\xi)-z\right)^{-N}
                                                + \left(\mu_-(x,\xi)-z\right)^{-N}]dxd\xi
                                                 \eeq
                                                 Now we shall complete  the proof of the proposition by contradiction.
                                                  Assume that there exists a sequence $\hbar_j$, 
                                                  $\lim_{j\rightarrow +\infty}\hbar_j = 0$,  such that $\hat A$ has no eigenvalues in  a neighborhood of $\pm i$. It follows that 
                                          $F_j(z):= F_{\hbar_j}(z)$, is holomorphic in
                                                                                a disc $\Omega_{r_1}$ for some $r_1 > 1$.
                                                                               Using Weyl-Ky Fan inequalities \cite{gokr},
                                                                               we can see that $F_j$ is a uniformly bounded sequence 
                                                                               of holomorphic functions in  $\Omega_{r_1}$. 
                                                                            Using    Montel's Theorem,
                                                 by taking  a  subsequence, we can assume that
                                                 $\lim_{j\rightarrow +\infty}F_j = F_\infty$
                                                  exists and is holomorphic in $\Omega_{r_1}$.
                                                  Then $F_{c\ell}$ has an holomorphic extension in $\Omega_{r_1}$.
                                                   Hence  we get a contradiction by computing
                                                   \beq
                                                   \lim_{s<1, s\rightarrow 1}\vert F_{c\ell}(is)\vert = +\infty
                                                   \eeq
                                                   \QED
                                                   We can apply the above result to improve a little bit a result of \cite{herowa}
                                                   \begin{corollary} Let us assume that $P(x^\prime)$
                                                    and $Q(x^{\prime\prime})$ are elliptic  polynomials in
                                                    $\mathbb R^{n^\prime}$ of degree $m^\prime$, 
                                                     respectively  in $\mathbb R^{n^{\prime\prime}}$ of degree     $m^{\prime\prime}$.
                          Then for every $n^\prime$, $n^{\prime\prime}$, $m^\prime\geq 2$,
                                                                                                           the  quadratic pencil
                                                                      $L(\lambda) = -\triangle_{x^\prime,x^{\prime\prime}}
                                                                       + Q^2(x^{\prime\prime}) + \left(P(x^\prime)-\lambda\right)^2$
                                                                        has an infinite number of eigenvalues
                                                                    \end{corollary}
                                                                    {\bf Proof}\\
                                                                   Let us remark that the self-adjoint operator 
                                                                    $K:=-\triangle_{x^{\prime\prime}}+Q^2(x^{\prime\prime})$
                                                                    has
                                                                    a basis, $\{\varphi_j\}$, of eigenfunctions in $L^2(\mathbb R^{n^\prime})$, with eigenvalues $\eta_j$ such that
                                                                     $\lim_{j\rightarrow +\infty}\eta_j = +\infty$. So the corollary is a consequence of Proposition \ref{loc}
                                                                     \QED
     We can get also  estimates on the number of eigenvalues
     for the pencil $L_\eta(\lambda)$, in every dimension,  for $\eta$ large enough.
      Let us introduce $N_\eta(R) = \#\{j\geq 1,\; \vert\lambda_j(\eta)\vert \leq R\}$.
      \begin{proposition}
      There exist $C>1$, $R_0>0$, $\eta_0>0$ such that,  for $R\geq R_0$, $\eta \geq \eta_0$
       we have
       \beq
       \frac{(\eta R)^{n(m+1)/m}}{C} \leq  N_\eta(R) \leq C(\eta R)^{n(m+1)/m}.
       \eeq
       \end{proposition}
       {\bf Sketch of Proof}\\
       We follow the same method as for proving Proposition \ref{estimate}. For convenience,
        we work in the semiclassical side. Le us denote
         $\tilde{N}_\hbar(R) = \#\{j\geq 1,\; \vert\mu_j(\hbar)\vert \leq R\}$.\\
         We prove first the upper bound. If $s_j(\hbar)$ denotes the eigenvalues of 
         $(\hat A^*\hat A)^{1/2}$,  spectral and semiclassical analysis \cite{ro2, ko}
          gives, that for some
          constants $K>0$, $\varepsilon_0 >0$, we have
          \beq\label{eq15}
          \sum_{j\geq 1} (s_j(\hbar)+t)^{-k} \leq K\hbar^{-n}t^{\theta-k}
          \eeq
          for $\hbar \leq  \varepsilon_0$, $t\geq 1$.  Then using Weyl-Ky Fan's inequality, 
           we get as  in Proposition\ref{estimate},
           \beq\label{eq16}
           \tilde{N}_\hbar(R) \leq K\hbar^{-n}R^{\theta}.
           \eeq
           For the lower bound,  we first remark that from   the trace formula (Theorem \ref{trace})
            we get for some $c_0 >0$,
             \beq\label{eq17}
             \sum_{j\geq 1}(\vert\mu_j(\hbar)\vert+t)^{-k} \geq c_0\hbar^{-n}t^{\theta-k}
             \eeq
            So, for $\hbar$ small enough, we have
            \beq
            \int_{1/2}^{+\infty}(r+t)^{-k}d\tilde N_\hbar(r) \geq c_0\hbar^{-n}t^{\theta-k}.
            \eeq
            Using the upper bound, we can choose $\gamma >0$ large enough, such that, 
             for $R\geq R_0$,  we have 
            \beq
            \int_{\gamma R}^{+\infty}r^{-k}d\tilde N_\hbar(r)  \leq 
            \frac{c_0}{2}\hbar^{-n}R^{\theta-k}, 
            \eeq
             which gives easily the lower bound
             \beq
             \tilde N_\hbar(\gamma R)  \geq  2^{-k-1}c_0\hbar^{-n}R^{\theta}.
             \eeq
             \QED
                   
  {\it Acknowledgement}. {\footnotesize The author thanks B. Helffer and Xue Ping Wang for their comments on this paper. }

\end{document}